# Electron Holography study of the local magnetic switching process in MTJs

E. Javon[1,2], C.Gatel[1,2], A. Masseboeuf[1,2], E.Snoeck[1,2,]

1 - CNRS; CEMES ; BP 94347, F-31055 Toulouse Cedex, France

2 - Université de Toulouse ; UPS ; CEMES ; BP 94347, F-31055 Toulouse Cedex, France

**Abstract**

We present an electron holography experiment enabling the local and quantitative study of magnetic properties in magnetic tunnel junction. The junction was fully characterized during the switching process and each possible magnetic configuration was highlighted with magnetic induction maps. No magnetic coupling was found between the two layers. We plot a local hysteresis loop that was compared with magnetometry measurement at the macroscopic scale confirming the validity of the local method.

**Introduction**

Different techniques may be used for the local study of magnetic properties in nanosystems like spin polarised scanning tunnelling microscopy, magnetic force microscopy or Lorentz microscopy. Electron holography (EH) allows a unique combination of the high spatial resolution of transmission electron microscopy (TEM) with the capability to quantitatively analyse local magnetic configurations [1, 2]. These EH capacities are particularly useful to record local magnetic map during switching processes and to plot the corresponding local hysteresis loop.

We used EH to investigate the magnetic configurations of an epitaxial Au/Co/Fe/MgO/FeV/MgO(001) magnetic tunnel junction (MTJ) grown by molecular beam epitaxy. This MTJ exhibits a giant tunnel magneto-resistance (TMR) of about 200% at RT [3]. The TMR properties appear when the magnetic configuration of the two FeV and Fe/Co magnetic layers switches from parallel to antiparallel arrangement. The aim of this article is to locally map the magnetic configurations of the MTJ during the magnetization reversal and to reconstruct the hysteresis loop from quantitative information obtained by EH.

**Experiments**

The study has been performed using a FEI Tecnai F-20 microscope (200 kV) equipped with a field emission gun and a Cs corrector. Holograms were acquired with the specimen in a field free region in the TEM column by turning off the main objective lens and using the first transfer lens of the Cs corrector like an objective lens [4]. Thin samples were prepared in cross-sectional geometry using the usual mechanical polishing then precision ion polishing method.

The in-situ magnetization was done by switching on the main objective lens with a slight excitation current in order to create a magnetic field of $\mu_0H$= 0.33 T along the optic axis of the microscope without distorting the resulting image. The sample was then progressively tilted in order to apply a controlled in plane magnetic field parallel to the layers. The axial component of the applied field has no influence on the reversal process because of the shape anisotropy of the sample and therefore can be neglected in our experiment. The coercive fields of the [Fe/Co] hard and FeV soft ferromagnetic layers are 18 mT and 2 mT respectively. Tilting from -13° to 13° corresponds to the application of an in plane magnetic field between ± 45 mT on the sample, which is thus sufficient to achieve the magnetic reversal processes of the whole MTJ and the investigation of the whole hysteresis loop. Electron holograms were then recorded with a varying in-plane magnetic field.

EH is an interferometric technique which gives access to the phase shift of an electron wave that has interacted with magnetic and electrostatic potentials. This phase shift is directly related to the electro-magnetic fields in the sample as described by the Aharonov-Bohm effect: $\Delta \Phi = C_E \int V(x, z)dz - \frac{e}{\hbar} \iint B_y(x, z)dxdz$ [1]. In that one dimensional expression, "x" is lying perpendicular to the electron beam (in our experiment "x" is in the plane of the sample perpendicular to the Co/Fe/MgO/FeV interfaces), and "z" is parallel to the electron beam. $C_E$ is an electron energy related constant and V is the electrostatic potential produced by the sample. "$B_y$" is the magnetic induction component perpendicular to both "x" and "z". The electrostatic contribution to the phase shift is directly proportional to the mean inner potential (MIP) of the sample and to its thickness if the MIP is constant along the electron wave path (as in our case). In order to isolate the magnetic component of the phase shift we acquired two holograms obtained with magnetization saturated in two opposite directions. We then have extracted a reference MIP phase image from holograms recorded at +13° and -13° which was then used to extract the magnetic contribution for each tilt angle : $\phi_{mag}(\theta) = \iint B_y(\theta)dxdz = -\frac{\hbar}{e}(\Delta \Phi(\theta) - MIP_{réf})$

**Results and discussion**

Two examples of the phase images corresponding to the magnetic contribution to the phase shift are displayed in Figure 1 (a). The map of the magnetic induction components ($B_x$ and $B_y$) in the two layers can be obtained calculating the derivatives: $\frac{\partial(\phi_{mag})}{\partial x} = B_y.t$ and $\frac{\partial(\phi_{mag})}{\partial y} = B_x.t$ (t being the sample thickness) and are reported in Figure 1 (b) and (c). They indicate that the magnetic induction is only running parallel to the layers ($B_x = 0$) and that for +9° tilt the magnetizations in the two Fe/Co and FeV layers are parallel while, for -1° tilt, there are antiparallel. We thus can get the magnetic configuration change of the MTJ for each step of the switching process.

In order to obtain quantitative results we have calculated the values of relative magnetization for each magnetic configuration. The same area has been selected in each "$B_y$" map and the values of all pixels of the "$B_y$" components have been summed to get: $I = \sum_{pixels} B_y t$. This sum is related to the magnetic induction along "y" and to the sample thickness. We then can use the MIP profile to get the shape of the sample thickness. It indicates an expected wedge shape of the sample along "x" with a linear slope of 0.95. The total intensity "I" does not give information about the real magnetization but enable us to compare the global magnetization of the junction for all the steps during the switching process. The evolution of "I" as a function of the tilt (i.e. the applied field) has been superimposed on the same plot than the hysteresis loop measured on the macroscopic sample by VSM (vibrating sample magnetometer). These two plots are reported in Figure 2. Experimental points from EH measurements match fairly well the VSM hysteresis loop. The coercive fields of the soft (FeV) and hard (Fe/Co) layers measured by EH agree with the ones measured by VSM. No dipolar coupling between the layers that could occur due to the thin sample preparation for TEM and EH studies are evidenced. Furthermore, the local configuration at -1° tilt (i.e. 3.5 mT) shows the appearance of dark contrast within the FeV layer corresponding to zero magnetisation (antiparallel configuration within the layer along the optical axis) or a magnetization oriented parallel to the optic axis after the switching started. It corresponds to the reversal of magnetic domain leading to the full reversal of the soft magnetic layer.

**Conclusion**

The agreement between measurements extracted from holography study and VSM shows that electron holography is a powerful tool to investigate the magnetic configurations of magnetic nanomaterials and nanosystems like MTJs. Hysteresis loop along any directions can be reconstructed from a series of maps and give quantitative values on the magnetic properties as coercive field. The main advantage of the technique is to couple the measurement and the image showing the magnetic configuration. This technique brings local information which can not be deduced from a simple hysteresis loop.

**Acknowledgement**

The authors acknowledge financial support from the European Union under the Framework 6 program under a contract for an Integrated Infrastructure Initiative. Reference 026019 ESTEEM and from the French ANR program "Spinchat" N°: BLAN07-1_ 188976. F. Bonell and S. Andrieu are acknowledged for providing the MTJ samples.

**References**

[1]: R.E.Dunin-Borkowski, M.R.McCartney, D.J.Smith, Electron Holography of Nanostructured Materials, Encyclopedia of nanoscience and nanotechnology, **10**, 1-59, 2003.

[2]: M.Heumann,T.Uhlig and J.Zweck, True Single Domain and Configuration-Assisted of Submicron Permalloy Dots observed by Electron Holography, Phys. Rev. Lett **94**, 077202, 2005.

[3]: F.Bonell, S.Andrieu and al., MgO Epitaxial Magnetic Tunnel Jonctions using FeV electrodes, Proceeding IMM, 2009.

[4]: E. Snoeck, P Hartel, H Müller, M Haider and P C Tiemeijer, Proceeding IMC Sapporo (2006)

**Figure captions**

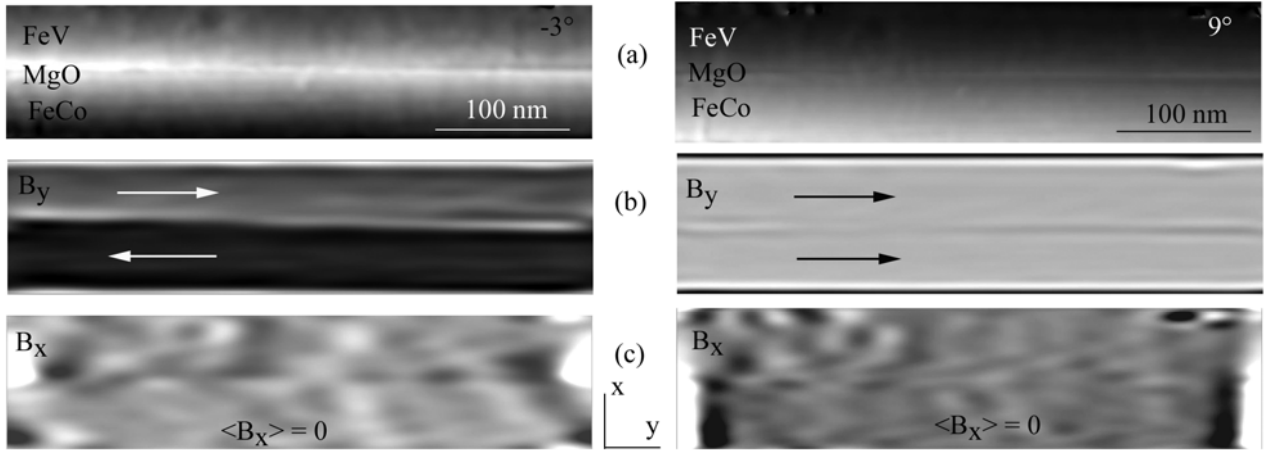

Figure 1: (a) Magnetic phase contribution to the phase shift. (b) "x" derivative of the magnetic phase images mapping the "y" components of the induction. (c) "y" derivative of the magnetic phase images mapping the "x" components of the induction.

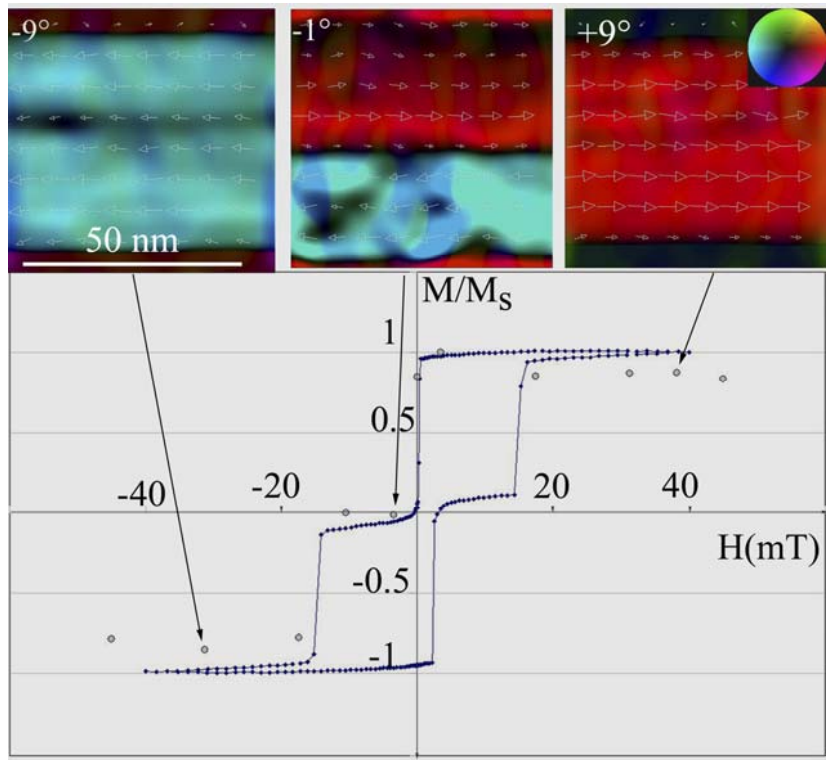

Figure 2: Superimposition of the normalized magnetization $M/M_S$ as a function of the applied magnetic field obtained by VSM (blue dots) and EH (pink dots). Corresponding vectorial map of the magnetization are given for 3 values of the tilt i.e. of the applied magnetic field (Inset top right: the color scale showing the direction and intensity of the magnetic induction)